# Dual-channel Transfer and Modulation of Optical Vortices in a Ladder-Type System

YAZHI SHAN AND YAN ZHANG*

*School of Physics, Northeast Normal University, Changchun 130024, China*

**Abstract:** We propose a dual-channel closed-loop structure within a symmetry-broken ladder-type three-level quantum system, where each channel incorporates three-wave mixing (TWM) processes. This system enables the transfer of optical vortices from either the strong control field or the weak input probe field to the TWM-generated fields in the channel. As a result, the two resultant output optical fields exhibit unique spatial distributions, such as crescent and petal shapes, which can be non-synchronously modulated via adjusting the intensity and frequency of the control field. The orbital angular momentum transfer dynamics and multi-angle spatial distribution modulation of the output fields show periodic and non-synchronous characteristics via adjusting the control field due to the optical interference. Thus, this quantum system has great potential as a tunable optical modulator, which holds promise for applications in information technologies involving multi-channel processing and asymmetric tasks, especially in advanced optical modulation and quantum state manipulation in quantum information processing.

## 1. Introduction

In the past few decades, the study of optical vortices and their manipulation has blossomed into a rapidly evolving frontier at the intersection of optics and information science. The novel properties of optical vortices find diverse applications [1-7]. Optical vortices are distinguished by their quantized orbital angular momentum (OAM) and phase singularity, presenting a helical wavefront and a central intensity null [8]. These features have been harnessed to advance technologies such as optical tweezers [1], where the OAM enables precise manipulation of microscopic particles, and various information processing applications [2-7]. Notably, encoding information in the OAM of vortex beams has made them indispensable in optical manipulation and trapping [9-11], high-precision metrology [12,13], and optical communications [14, 15]. For example, the combination of various OAM states offers the potential to increase data transmission capacities significantly. Moreover, the dynamics of OAM transfer with the optical frequency conversions in nonlinear optical processes, including second- or higher-order nonlinear interactions, has been extensively investigated, enabling novel functionalities [16-19].

Recent research has made remarkable progress in elucidating the mechanisms of optical vortex transfer, particularly through the use of quantum systems. The transfer of OAM from a strong control field to a weak probe field has been demonstrated during light storage and retrieval processes [20, 21], the propagation of OAM via four-wave mixing interactions [22-24], the noise-induced coherence effect on the OAM transfer [25], and the transfer between two weak fields mediated by ground-state coherence [26]. One significant development is the enhanced optical vortex efficiency achieved in nonlinear schemes based on symmetry-broken quantum systems [27]. Unlike natural atoms, which typically possess inversion symmetry in their potential energy and thus impede three-wave mixing (TWM) processes, symmetry-broken quantum systems, such as ladder-type three-level configurations, can form a closed-loop structure [28-33]. This structure allows for the generation of TWM beams by circumventing the conventional electric dipole selection rule. The realization of such symmetry-broken quantum systems often relies on carefully engineered materials, including artificial atoms with tunable energy levels, semiconductor quantum wells (or dots), and nanomaterials with optimized parameters.

For multi-task scenarios involving OAM, multi-channel transfer and modulation capabilities are essential. In this paper, we propose a dual-channel closed-loop structure within a symmetry-broken ladder-type three-level quantum system. This structure enables the transfer of optical vortices from the control field or probe fields to the TWM-generated fields through two independent TWM processes in separate channels. Under suitable conditions, optical interference between the TWM-generated fields and the incident probe fields of the same frequency gives rise to special spatial distributions of the resultant output fields. The interference effect allows the control field to spatially modulate the resultant output fields. By adjusting the detuning and OAM of the control field, we can achieve non-synchronous modulation on the structured light beams. These findings may have potential applications in asymmetric modulation in quantum information processing tasks.

## 2. Model and Equations

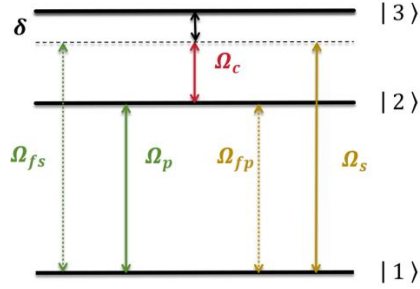

Fig. 1 Schematic diagram of the dual-channel closed-loop structure in a symmetry-broken ladder-type three-level quantum system.

Here, we propose a symmetry-broken ladder-type three-level quantum system, as depicted in Fig. 1. This system contains a ground level $|1\rangle$ and two excited states $|2\rangle$ and $|3\rangle$. The transition $|3\rangle \rightarrow |2\rangle$ is driven by a strong control field with Rabi frequency $\Omega_c$ and central frequency $\omega_c$. A weak field with Rabi frequency $\Omega_p$ and central frequency $\omega_p$, probes the transition $|2\rangle \rightarrow |1\rangle$, while a weak field with Rabi frequency $\Omega_s$ and central frequency $\omega_s$, probes the transition $|3\rangle \rightarrow |1\rangle$. These fields form two closed-loop configurations, namely the transition pathway $|1\rangle \rightarrow |2\rangle \rightarrow |3\rangle \rightarrow |1\rangle$ and its reverse pathway $|1\rangle \rightarrow |3\rangle \rightarrow |2\rangle \rightarrow |1\rangle$. The transition $|3\rangle \rightarrow |1\rangle$ is allowed to generate new fields with frequencies $\omega_{fs} = \omega_p + \omega_c = \omega_s$ and $\omega_{fp} = \omega_s - \omega_c = \omega_p$, respectively. To ensure the energy matching condition, the detunings must satisfy $\delta_{fs} = \delta_p + \delta_c$ and $\delta_{fp} = \delta_s - \delta_c$. We assume $\delta = \delta_c = \delta_s$, and $\delta_p = 0$, where $\delta_c$ is the detuning of the control field, and $\delta_{fp}$, $\delta_{fs}$, $\delta_p$ and $\delta_s$ are the detunings of the TWM-generated fields and the probe fields, respectively.

Under the rotating-wave approximation, the interaction Hamiltonian can be expressed a ($\hbar = 1$):

$$H_I = \delta|3\rangle\langle 3| - \frac{1}{2}(\Omega_p|2\rangle\langle 1| + \Omega_{fs}|3\rangle\langle 1| + \Omega_{fp}|2\rangle\langle 1| + \Omega_s|3\rangle\langle 1| + \Omega_c|3\rangle\langle 2| + H.C.),$$
(1)

where $\Omega_{fs}$ and $\Omega_{fp}$ are the Rabi frequency of the TWM-generated fields of the sum-frequency process and the difference-frequency process, respectively.

We further assume that both probe fields are much weaker than the control field, and we can treat the contribution of the probe fields as a perturbation in the equations of motion of the density matrix elements. Then the elements $\rho_{31}$ and $\rho_{21}$ are given by

$$\dot{\rho}_{31} = -(\gamma_{31} + i\delta)\rho_{31} + \frac{i}{2}(\Omega_s + \Omega_{fs}) + \frac{i}{2}\Omega_c \rho_{21}, \tag{2}$$

$$\dot{\rho}_{21} = -\gamma_{21}\rho_{21} + \frac{i}{2}(\Omega_p + \Omega_{fp}) + \frac{i}{2}\Omega_c^* \rho_{31}. \tag{3}$$

Here, $\gamma_{31}$ and $\gamma_{21}$ are coherent decay rates of relevant transitions. By solving the steady-state solutions of these equations, we obtain

$$\rho_{31}^{(1)} = \frac{i(\Omega_s + \Omega_{fs})\gamma_{21}}{2Y}, \tag{4}$$

$$\rho_{21}^{(1)} = \frac{i(\Omega_p + \Omega_{fp})(\gamma_{31} + i\delta)}{2Y}, \tag{5}$$

$$\rho_{31}^{(2)} = \frac{i^2(\Omega_s + \Omega_{fs})\Omega_c}{4Y}, \tag{6}$$

$$\rho_{21}^{(2)} = \frac{i^2(\Omega_p + \Omega_{fp})\Omega_c^*}{4Y}, \tag{7}$$

with $Y = \gamma_{21}(\gamma_{31} + i\delta) + |\Omega_c|^2/4$. Eqs. (4) and (5) typically represent linear processes reflecting the absorption and dispersion of two probe fields, respectively. Eqs. (6) and (7) characterize two second-order nonlinear TWM processes, respectively.

Under the slowly varying envelope approximation, Maxwell's equations of the probe fields and the TWM-generated fields can be expressed as

$$\frac{\partial \Omega_s}{\partial z} = \frac{i\alpha_s \gamma_{31}}{2L}\left(\frac{i\Omega_s \gamma_{21}}{2Y} + \frac{i^2 \Omega_{fp}\Omega_c}{4Y}\right), \tag{8}$$

$$\frac{\partial \Omega_{fp}}{\partial z} = \frac{i\alpha_p \gamma_{21}}{2L}\left(\frac{i\Omega_{fp}(\gamma_{31} + i\delta)}{2Y} + \frac{i^2 \Omega_s \Omega_c^*}{4Y}\right), \tag{9}$$

$$\frac{\partial \Omega_p}{\partial z} = \frac{i\alpha_p \gamma_{21}}{2L}\left(\frac{i\Omega_p(\gamma_{31} + i\delta)}{2Y} + \frac{i^2 \Omega_{fs}\Omega_c^*}{4Y}\right), \tag{10}$$

$$\frac{\partial \Omega_{fs}}{\partial z} = \frac{i\alpha_s \gamma_{31}}{2L}\left(\frac{i\Omega_{fs}\gamma_{21}}{2Y} + \frac{i^2 \Omega_p \Omega_c}{4Y}\right).$$

(11)

Here, $\alpha_s$ and $\alpha_p$ represent optical depths of the two probe fields, and $L$ represents the length of the medium. We ignore the diffraction term containing the transverse derivatives $\left(2k_{p(fp)}\right)^{-1} \nabla_\perp^2 \Omega_{p(fp)}$ and $\left(2k_{s(fs)}\right)^{-1} \nabla_\perp^2 \Omega_{s(fs)}$, where $k_{p(fp)}$ and $k_{s(fs)}$ are the wave vectors of the relevant fields. This causes the change of the phase due to the diffraction term after the beams pass through the medium is approximately $L/2k\omega^2$, which can be neglected, when the propagation distance is much smaller than the Rayleigh range of the fields.

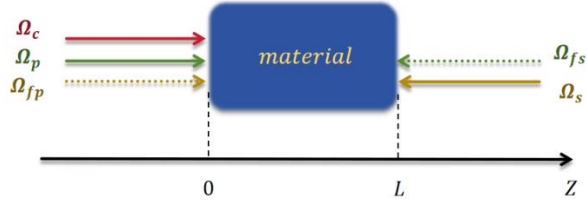

Fig. 2 Schematic diagram of the interaction between each optical field and the material medium

As shown in Fig. 2, the initial conditions of the probe fields are set as follows: $\Omega_p(z=0) = \Omega_{p0}$ for the left side, and $\Omega_s(z=L) = \Omega_{s0}$ for the right side. To facilitate the derivation of the propagation equations and enhance the clarity of the results, we assume that $\alpha_u \gamma_{31} = \alpha_d \gamma_{21} = d$, then the solutions to Eqs. (8-11) are:

$$\Omega_{s(z)} = \Omega_{s0} \left[\cos\frac{\beta dz}{8YL} - \frac{(\gamma_{21} - \gamma_{31} - i\delta)}{\beta} \sin\frac{\beta dz}{8YL}\right] \exp\left[-\frac{dz}{8YL}(i\delta + \gamma_{31} + \gamma_{21})\right],$$
(12)

$$\Omega_{fp(z)} = -\frac{i\Omega_c^* \Omega_{s0}}{\beta} \sin\frac{\beta dz}{8YL} \exp\left[-\frac{dz}{8YL}(i\delta + \gamma_{31} + \gamma_{21})\right],$$
(13)

$$\Omega_{fs(z)} = -\frac{i\Omega_c \Omega_{p0}}{\beta} \sin\frac{\beta dz}{8YL} \exp\left[-\frac{dz}{8YL}(i\delta + \gamma_{31} + \gamma_{21})\right],$$
(14)

$$\Omega_{p(z)} = \Omega_{p0} \left[\cos\frac{\beta dz}{8YL} - \frac{(\gamma_{31} + i\delta - \gamma_{21})}{\beta} \sin\frac{\beta dz}{8YL}\right] \exp\left[-\frac{dz}{8YL}(i\delta + \gamma_{31} + \gamma_{21})\right],$$
(15)

where $\beta = \sqrt{|\Omega_c|^2 - (i\delta + \gamma_{31} - \gamma_{21})^2}$. Since the TWM-generated fields and the probe fields with the same frequency are indistinguishable, the resultant output fields $\Omega_d$ and $\Omega_u$ are the coherent superposition of them, i.e., $\Omega_{d(z)} = \Omega_{p(z)} + \Omega_{fp(L-z)}$ and $\Omega_{u(z)} = \Omega_{s(L-z)} + \Omega_{fs(z)}$, respectively.

The vortex beams are used as Laguerre-Gaussian (L-G) beams and are described as

$$\Omega_j = \varepsilon_j \left(\frac{r}{w}\right)^{|l_j|} e^{-\left(\frac{r}{w}\right)^2} e^{il_j\theta},$$
(16)

where $l_j$ is the topological charge (TC), i.e. the OAM number, while $\theta$, $w$, $\varepsilon_j$ and $r$ are the azimuthal angle, the waist parameter, the amplitude of the beam, and a cylindrical radius, respectively.

## 3. Results and Discussions

First, the control field $\Omega_c$, modeled as an L-G beam, serves as the source of optical vortices, playing a pivotal role in the transfer and modulation processes. Figure 3 showcases the intensity and phase profiles of the TWM-generated difference-frequency field $\Omega_{fp}$ at $z = 0$ and the TWM-generated sum-frequency field $\Omega_{fs}$ at $z = L$ for different TCs $l_c$ of the control field $\Omega_c$. From Fig. 3(a) and (c), it shows that the phase changes of the fields $\Omega_{fp}$ and $\Omega_{fs}$ are periodic, and their phase changes are opposite, which means that the OAMs carried by them are in opposite directions. In addition, The degree of phase change is intricately linked to $|l_c|$. As shown in Fig. 3(b) and (d), in terms of intensity, the radius of the annular distribution of $\Omega_{fp}$ and $\Omega_{fs}$ exhibits a synchronous increase with $|l_c|$, which is a characteristic trait of vortex beams. This behavior can be theoretically explained by Eqs. (13) and (14). The sum-frequency field $\Omega_{fs}$ contains the Rabi frequency of the control field $\Omega_c$, resulting in the same TC as $\Omega_c$, while the difference-frequency field $\Omega_{fp}$ contains the conjugate term $\Omega_c^*$, leading to an opposite TC. These results firmly demonstrate the successful transfer of optical vortices from the control field $\Omega_c$ to the TWM-generated fields $\Omega_{fp}$ and $\Omega_{fs}$ through the dual-channel mechanism proposed. Additionally, the difference between phases of $\Omega_{fp}$ and $\Omega_{fs}$ highlight the feasibility of non-synchronous spatial modulation of the resultant output fields by the control field $\Omega_c$.

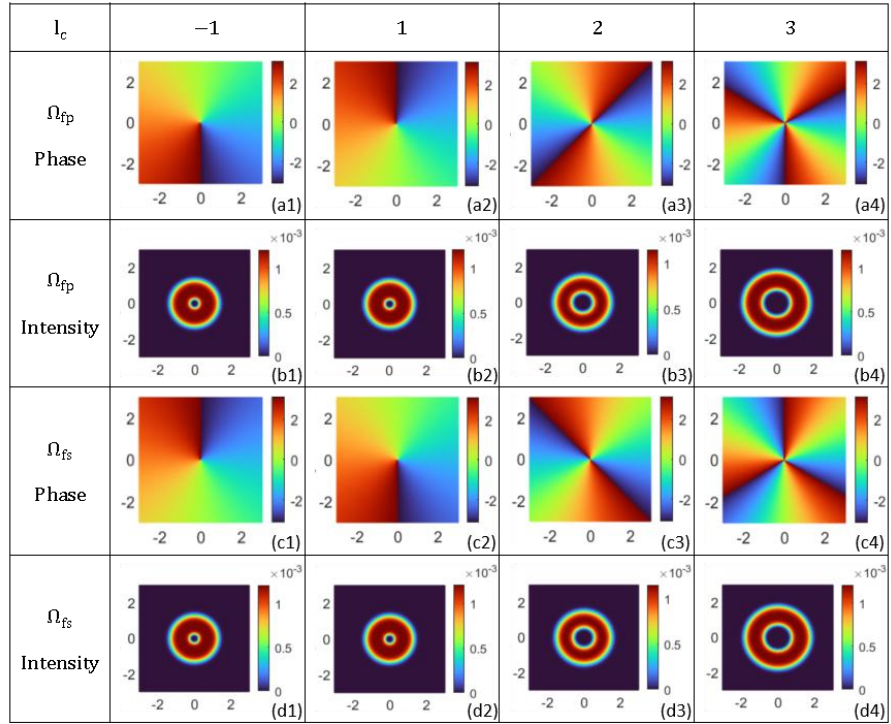

Fig. 3 Phase and intensity profiles of the TWM-generated fields $\Omega_{fp}$ at $z = 0$ in (a-b) and $\Omega_{fs}$ at $z = L$ in (c-d) for different TC $l_c$ of the control field. Horizontal and vertical axes are normalized to the beam waist parameter. Other parameters are $l_s = l_p = 0$, $\gamma = 10$ MHz, $\Omega_c = 4\gamma$, $\delta = 0$, $\gamma_{31} = \gamma$, $\gamma_{21} = 0.05\gamma$, $\Omega_{p0} = \Omega_{s0} = 0.1\gamma_{21}$, and $d = 100$.

Figure 4 presents the intensity and phase profiles of the resultant output fields $\Omega_d$ at $z = 0$ and $\Omega_u$ at $z = L$ for different detuning values $\delta$. We assume that all incident fields carry optical vortices with the same TC, specifically $l_c = l_s = l_p = 1$, where $l_s$ and $l_p$ are the TCs of the two probe fields $\Omega_p$ and $\Omega_s$. It is clear from Fig. 4(a) and (c) that the increasing of $|\delta|$ can suppressed the phase twists of the resultant fields $\Omega_d$ and $\Omega_u$ to a certain degree. This phenomenon is attributed to the fact that detuning weakens the light-matter coupling, thereby reducing the OAM-induced spatial interference. The impact of detuning $\delta$ on the intensity distributions of the resultant output fields $\Omega_d$ and $\Omega_u$ is also significant. As shown in Fig. 4(b) and (d), the intensity distributions exhibit crescent-shaped structures due to OAM-induced spatial interference with $l_c = l_s = l_p = 1$. When the detuning $\delta$ is zero, the notches of the crescent-shaped intensity distributions of $\Omega_d$ and $\Omega_u$ face exactly opposite directions. As $\delta$ increases (from $-9\gamma$ to $9\gamma$), their notches start to rotate in opposite directions, respectively. In addition, the intensities of both fields decrease with increasing $|\delta|$. This indicates that adjusting the frequency of the control field enables independent dynamic non-synchronous modulations of the spatial distributions of the resultant output optical fields in the two channels, with the modulation directions being opposite to each other.

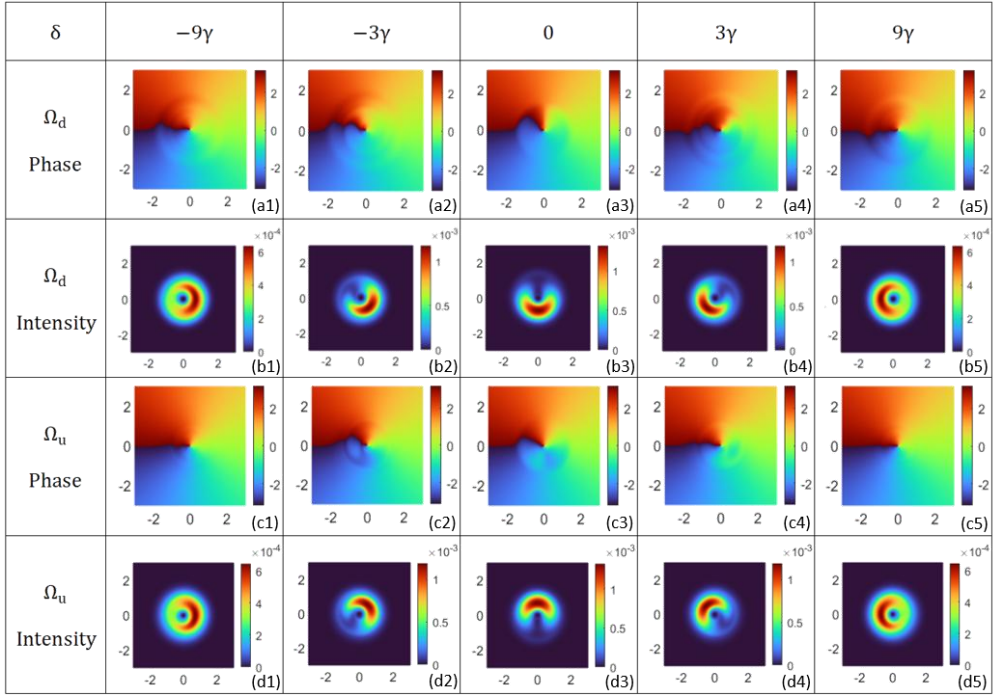

Fig. 4 Phase and intensity profiles of the resultant output fields $\Omega_d$ at $z = 0$ in (a, b) and $\Omega_u$ at $z = L$ in (c, d) for different detuning $\delta$. Horizontal and vertical axes are normalized to the beam waist parameter. Here $l_c = l_s = l_p = 1$, and other parameters are the same as in Fig. 3.

To gain a more comprehensive understanding of the effect of detuning $\delta$ on the intensity distribution of the resultant output fields $\Omega_d$ and $\Omega_u$, figure 5 shows the the evolution curves of the intensity of field $\Omega_d$ at $z = 0$ and $\Omega_u$ at $z = L$ with respect to the azimuthal angle $\theta$ of the control field for different detuning values. The intensity of field $\Omega_d$ ($\Omega_u$) oscillates periodically with the azimuthal angle $\theta$. Notably, increasing the detuning $\delta$ both the oscillation amplitude and the peak-valley difference of the spatial intensities of two fields. This indicates that the large detuning suppresses the intensity of resultant output fields and the interference effect. In addition, the intensity distribution exhibits a shift along angle $\theta$ with increasing $\delta$, means the rotation of the fields in the space. However, the peak position of field $\Omega_d$ always coincide with the valley position of of field $\Omega_u$ with $\delta = 0$; the shifts of of fields $\Omega_d$ and $\Omega_u$ are opposite along azimuthal angle $\theta$. These results from Figure 5 further confirm the opposite-direction rotation of spatial distributions of $\Omega_d$ and $\Omega_u$.

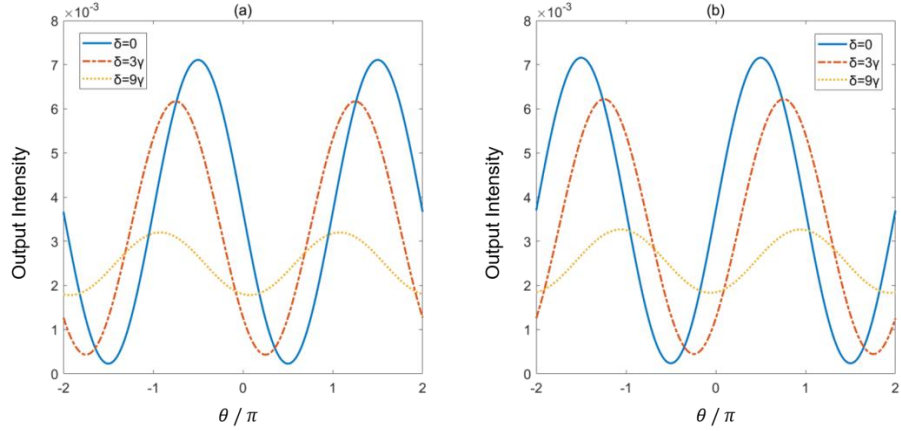

Fig. 5 Intensity of the resultant output fields $\Omega_d$ at $z = 0$ in (a) and $\Omega_u$ at $z = L$ in (b) versus the azimuthal angle $\theta$ of the control field for different detuning $\delta$. Here $l_c = l_s = l_p = 1$, and other parameters are the same as in Fig. 3.

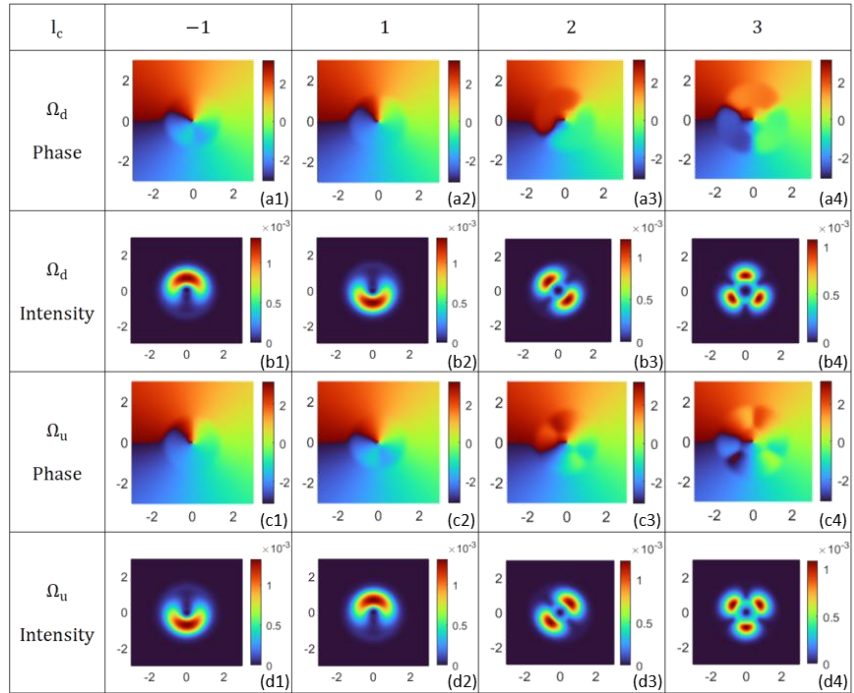

Fig. 6 Phase and intensity profiles of the resultant output fields $\Omega_d$ at $z = 0$ in (a, b) and $\Omega_u$ at $z = L$ in (c, d) for different TC of the control field. Both horizontal and vertical axes are normalized to the beam waist parameter. The TC satisfies $l_s = l_p = 1$, other parameters are the same as in Fig. 3.

Finally, we examine the intensity and phase profiles of the resultant output fields when the TC of the control field $\Omega_c$ is different from that of the probe field. Here, we assume $l_p = l_s = 1$ and vary $l_c$. When $\delta = 0$, both phase distributions of two resulatant fields exhibit a certain degree of distortion due to the interference. From Fig. 6(b) and (d), the intensity distributions of fields $\Omega_d$ and $\Omega_u$ exhibit petal-shaped structures. Specifically, the intensities of these two fields exhibit multiple periodic variation along the azimuthal angle $\theta$. When $l_p = l_s = 1$, the number of petals (i.e. the number of periods of spatial within a full spatial angular rotation corresponds to $|l_c|$. Notably, the orientations of the petal-shaped structures of fileds $\Omega_d$ and $\Omega_u$ are diametrically opposed. This is due to the presence of two TWM process. The difference-frequency field $\Omega_{fp}$ and the sum-frequency field $\Omega_{fs}$ acquire the same magnitude of OAM as the control field $\Omega_c$, but their OAM directions are opposite. Consequently, the TC of the control field also exerts an influence on the resultant output optical field formed by the interference between the probe field and the corresponding TWM-generated field. It affects the spatial distributions of the two resultant output fields, and these two spatial fractals are always opposite to each other.

## 4. Conclusion

In summary, we have investigated the dual-channel orbital angular momentum (OAM) transfer and the spatial modulation of optical vortex beams within a symmetry-broken ladder-type three-level quantum system. Through the utilization of a control field and two probe fields, this system is capable of supporting a sum-frequency and a difference-frequency TWM process concurrently. Our findings demonstrate that optical vortices can be transferred from the control field to the two TWM-generated fields. The two TWM-generated fields can acquire the same magnitude of OAM from the control field but with opposite directions. Owing to this distinct OAM-induced phase modulation, the spatial distribution of the dual-channel resultant output fields can be non-synchronously modulated by adjusting the frequency and TC of the control field. Specifically, when the probe fields carry OAMs, crescent-shaped or petal-shaped intensity distributions of the two resultant output fields can emerge. The structures of these dual-channel optical vortices can rotate counterclockwise by adjusting the control field. The control field assumes different roles in these two processes, resulting in non-synchronous responses of the TWM-generated fields to variations in the control field's parameters. Such non-synchronous modulation of dual-channel structured light beams holds significance in the design of certain nonlinear optical devices and has potential applications in OAM-based optical communication and optical information processing.